\documentclass[aps,pre,10pt,twocolumn,superscriptaddress,balancelastpage,showpacs,reprint]{revtex4-2}
\usepackage[english]{babel}
\addto{\captionsenglish}{}
 
\usepackage[utf8]{inputenc}
\usepackage[colorinlistoftodos, color=green!40, prependcaption]{todonotes}
\usepackage{centernot}
\usepackage{bm}
\usepackage{graphicx}
\usepackage{siunitx}
\usepackage{amsmath}
\usepackage{times}
\usepackage{amssymb}
\usepackage{mathrsfs}
\usepackage{color}
\usepackage{url}
\usepackage{version}
\usepackage[pdftex,colorlinks=true,pdfstartview=FitV,linkcolor= linkcolor,citecolor= linkcolor,urlcolor= linkcolor,hyperindex=true,hyperfigures=false]{hyperref}
\usepackage{balance}
\usepackage{lastpage}
    
\definecolor{linkcolor}{rgb}{0,0,0.6} 

\newcommand{\qqq}{\end{eqnarray}}

\newcommand{\pp}{{\bm p}}

\newcommand{\rr}{{\bm r}}
\newcommand{\RR}{{\bm R}}

\newcommand{\uu}{{\bm u}}
\newcommand{\UU}{{\bm U}}
\newcommand{\FF}{{\bm F}}

\begin{document}

\title{Collective motion in a sheet of microswimmers}

\author{D\'ora B\'ardfalvy}
\affiliation{Division of Physical Chemistry, Lund University, P.O. Box 124, S-221 00 Lund, Sweden}

\author{Viktor \v{S}kult\'ety}
\affiliation{SUPA, School of Physics and Astronomy, The University of Edinburgh, James Clerk Maxwell Building, Peter Guthrie Tait Road, Edinburgh, EH9 3FD, United Kingdom}
\affiliation{Neuro-X Institute, École Polytechnique Fédérale de Lausanne, Geneva, Switzerland}
\affiliation{Department of Radiology and Medical Informatics, University of Geneva (UNIGE), Geneva, Switzerland}

\author{Cesare Nardini}
\affiliation{Service de Physique de l'\'Etat Condens\'e, CNRS UMR 3680, CEA-Saclay, 91191 Gif-sur-Yvette, France}
\affiliation{Sorbonne Universit\'e, CNRS, Laboratoire de Physique Th\'eorique de la Mati\`ere Condens\'ee, LPTMC, F-75005 Paris, France.}

\author{Alexander Morozov}
\affiliation{SUPA, School of Physics and Astronomy, The University of Edinburgh, James Clerk Maxwell Building, Peter Guthrie Tait Road, Edinburgh, EH9 3FD, United Kingdom}

\author{Joakim Stenhammar}
\email{joakim.stenhammar@fkem1.lu.se}
\affiliation{Division of Physical Chemistry, Lund University, P.O. Box 124, S-221 00 Lund, Sweden}

\date{\today} 

\begin{abstract}
\textbf{Abstract.} Self-propelled particles such as bacteria or algae swimming through a fluid are non-equilibrium systems where particle motility breaks microscopic detailed balance, often resulting in large-scale collective motion. Previous theoretical work has identified long-ranged hydrodynamic interactions as the driver of collective motion in unbounded suspensions of rear-actuated (``pusher'') microswimmers. In contrast, most experimental studies of collective motion in microswimmer suspensions have been carried out in restricted geometries where both the swimmers' motion and their long-range flow fields become altered due to the proximity of a boundary. Here, we study numerically a minimal model of microswimmers in such a restricted geometry, where the particles move in the midplane between two no-slip walls. For pushers, we demonstrate collective motion with short-ranged order, in contrast with the long-ranged flows observed in unbounded systems. For front-actuated (``puller'') microswimmers, we discover a long-wavelength density instability resulting in the formation of dense microswimmer clusters. Both types of collective motion are fundamentally different from their previously studied counterparts in unbounded domains. Our results show that this difference is dictated by the geometrical restriction of the swimmers' motion, while hydrodynamic screening due to the presence of a wall is subdominant in determining the suspension's collective state.
\end{abstract}


\maketitle

\begin{figure}[b]
\center
\includegraphics[width=8.5cm,trim={0 0 7.5cm 13.1cm},clip]{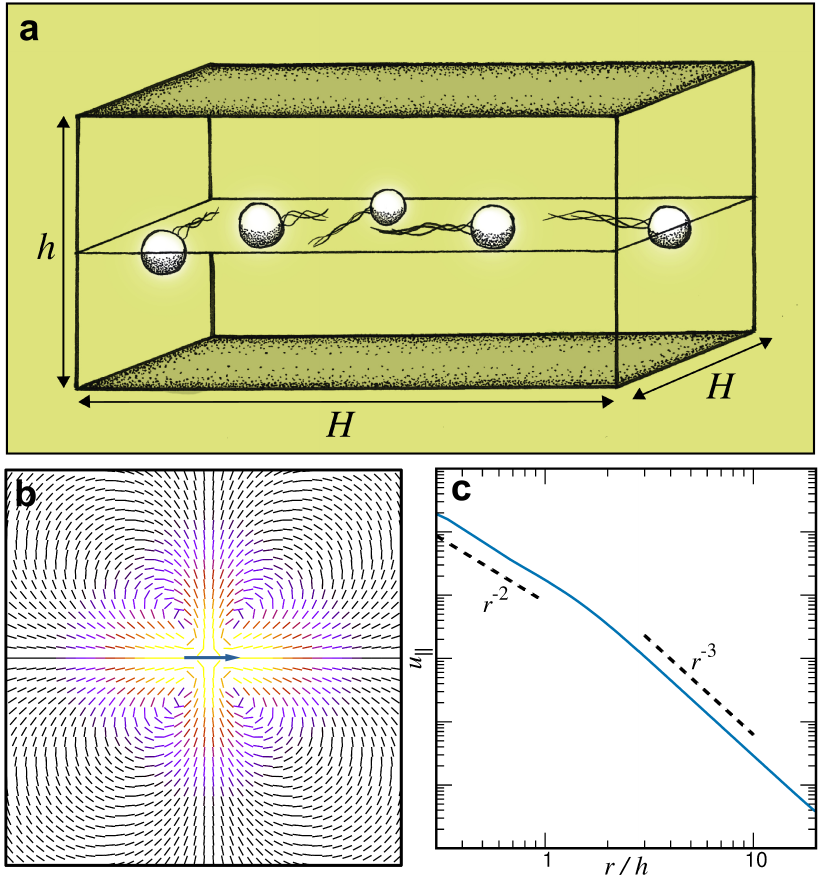}
\caption{\textbf{A single microswimmer between two no-slip walls. }(a) Schematic representation of the model system: the microswimmers are restricted to move in the mid-plane of a box of lateral dimensions $H \times H$ and confined by two no-slip walls separated by a height $h$. (b) Numerically calculated in-plane flow field of a confined pusher dipole; due to the confinement, the far-field flow has a quadrupolar symmetry. Vector colours indicate the relative magnitude of the flow on a logarithmic scale from black to yellow. (c) Magnitude $u_{\parallel}$ of the angle-averaged, in-plane flow field $\uu_{\parallel}$ in (b): for $r \gg h$, the flow field decays as $r^{-3}$ instead of $r^{-2}$. }\label{fig:1}
\end{figure}    

\section*{Introduction} \label{sec:intro}
Large-scale collective motion is a generic feature of biological and synthetic active matter systems across lengthscales. On the microscopic scale, one of the most well-studied systems is a suspension of flagellated microswimmers such as bacteria, algae or spermatozoa~\cite{Lauga1,Cates1,Yeomans1}. According to their swimming mechanism, one can differentiate between pusher- and puller-type microswimmers, where the former are rear-actuated and expel fluid along their main axis, while the latter are front-actuated and expel fluid perpendicular to this axis. The interactions in 3-dimensional ($3d$), unbounded microswimmer suspensions are often dominated by long-ranged hydrodynamic interactions (HIs)~\cite{Lauga1}, which for sufficiently high densities lead to hydrodynamically induced collective motion, so-called ``bacterial turbulence''~\cite{Creppy1,Wensink1,Hohenegger1,Koch:AnnuRev:2011}, a phenomenon absent in puller suspensions~\cite{Leptos1,Saintillan1}. This experimentally observed behaviour is in qualitative accordance with kinetic theories predicting a long-wavelength hydrodynamic instability above a threshold pusher density $n^c$; in unbounded, $3d$ pusher suspensions this threshold density, $n_{3d}^c$, is given by~\citep{Saintillan4,Skultety1}
  \begin{equation}\label{eq:nc_3d}
      n_{3d}^c = \frac{5\lambda}{B\kappa}, 
  \end{equation}
where $\lambda$ denotes the frequency of random bacterial reorientations (``tumbling''), $\kappa$ the dipolar strength, and $B$ is the Bretherton parameter describing the swimmer's shape, where $B = 0$ corresponds to a sphere and $B = 1$ to a needle-like particle~\citep{Kim2005}. In their natural habitats, many microswimmers however reside in geometrically restricted environments, such as microorganisms living in porous soil~\cite{Or-2007} or spermatozoa swimming in the reproductive tract~\cite{Suarez-2005}. Moreover, most laboratory experiments on biological microswimmers, including those where collective motion is observed, have been performed in quasi-$2d$ geometries such as near solid surfaces, in thin films, or in microfluidic devices~\cite{Polin:PRL:2019,Riedel1,Creppy1,Dombrowski1,Sokolov1,Wu1,Kantsler-2013,GuastoPRL,Pepper-2010}. Thus, direct comparisons between experiments performed in confined geometries and predictions from kinetic theories of unbounded microswimmer suspensions is not straightforward, due to the long-ranged nature of hydrodynamic swimmer-swimmer and swimmer-wall interactions. This subtlety is further confirmed by experimental measurements of flow fields around confined microswimmers~\cite{Polin:PRL:2019} which show that these strongly depend on the details of the swimming mechanism, swimmer orientation, and nature of the confinement. 

In unbounded suspensions, the transition to active turbulence is driven by mutual reorientation due to the long-ranged dipolar flow fields of pusher microswimmers, decaying as $r^{-2}$, where $r$ is the swimmer-swimmer separation~\cite{Lauga1,Baskaran1,Leoni1,Woodhouse1}. To leading order in $r$, the flow field at a point $\rr$ due to a swimmer with orientation $\pp$ placed at the origin reads
\begin{equation}\label{eq:u_dipole}
\uu_d (\rr) = \frac{\kappa}{8\pi} \left[ \frac{3(\pp\cdot\rr)^2}{r^2} - 1 \right] \frac{\rr}{r^3},
\end{equation}
where all the details of the microswimmer itself enter solely \emph{via} the dipolar strength $\kappa$. 

As was first shown by Liron and Mochon~\citep{Liron1}, confinement of a Stokeslet (\emph{i.e.}, a force monopole) between two no-slip walls separated by a distance $h$ (Fig.~\ref{fig:1}a) introduces a hydrodynamic screening in the direction parallel to the walls, leading to a faster power-law decay of the flow field. The corresponding flow field for a force dipole can then be obtained by differentiation of the Stokeslet expressions~\citep{Liron1,Daddi-Moussa-Ider_2018,Mathijssen:JFM:2016}, yielding a far-field flow which has the symmetry of a source quadrupole flow but decays as $r^{-3}$ for $r \gg h$~\citep{Liron1,Polin:PRL:2019}. In Fig.~\ref{fig:1}b and c, we show the corresponding flow field obtained numerically from a lattice Boltzmann simulation, as described further in the Methods section. In strongly confined systems, where $h$ is comparable to the microswimmer size, this screening suppresses bacterial turbulence and the dynamics are instead characterised by qualitatively different forms of collective motion dictated by the effective friction between the swimmer and the confining wall~\citep{Bartolo:PRL:2013,Saintillan:PRE:2014}. In the opposite limit, corresponding to $h \rightarrow \infty$, the microswimmer retains its bulk form~\eqref{eq:u_dipole}, and the system thus corresponds to a $2d$ sheet of microswimmers embedded in a $3d$ fluid. In spite of the absence of hydrodynamic screening in this limit, in a recent study~\citep{Skultety:JFM:2024} we showed that this restriction of the swimmer dynamics to a lower-dimensional space than that of the flow field induces a novel set of hydrodynamic instabilities that are qualitatively different from the fully $3d$ case. In that study, we analysed the stability of a homogeneous isotropic sheet of microswimmers described through mean-field kinetic theory, similar to the derivation previously carried out for $3d$ bulk suspensions leading to the instability criterion in Eq.~\eqref{eq:nc_3d}~\citep{Saintillan4,Skultety1}. For a sheet of pushers, we showed that the long-wavelength instability in the orientation field leading to active turbulence is transformed into a short-wavelength one, occuring at a lengthscale comparable to the typical swimmer-swimmer separation. Additionally, we showed that a $2d$ sheet of \emph{pullers} embedded in a $3d$ fluid exhibits a novel instability in the microswimmer density field. Unlike the pusher instability, this instability occurs on the scale of the system size and is enabled by the fact that the in-plane component $\uu_{\parallel} = (u_x,u_y)$ of $\uu_d$ in Eq.~\eqref{eq:u_dipole} is effectively compressible (\emph{i.e.} $\nabla_{2d} \cdot \uu_{\parallel} \neq 0$) even when the $3d$ fluid is incompressible, making pullers act as effective in-plane sinks and pushers as sources. This previously neglected effect is likely responsible for the clustering of active particles near boundaries observed previously in experiments~\citep{Stone:PNAS:2018} and simulations~\citep{Adhikari:PRL:2016} where the particle dynamics were restricted to a lower dimension than the swimmer flow field due to the presence of a fluid or a solid boundary. In equilibrium systems, this type of restriction has been shown to induce a substantial increase in the collective diffusion coefficient for Brownian colloids compared to the bulk case~\citep{Oettel:SoftMatter:2014,Oettel:JPCM:2015,Oettel:PRE:2017}. The analytic results in~\citep{Skultety:JFM:2024} focussed on the linear instability of the mean-field orientation and density fields, and therefore did not characterise the ensuing, nonlinear steady states, which requires numerical simulations; this topic is instead the focus of the present study. 

The effects of confinement on collective motion in microswimmer suspensions have previously been addressed mainly using particle-based simulations of both dipolar microswimmers~\citep{Graham:PRL:2005,Graham:JPCM:2009,Menzel:MolPhys:2018} and ``squirmers'' that swim through tangentially imposed flow field along their bodies~\cite{Alarcon:2013,Zantop_SoftMatter_2020, Liverpool:PRE:2017, Gompper:SoftMatter:2018, Qi:CommunPhys:2022,Zottl:PRL:2014}. These studies demonstrated a range of collective phenomena depending on the properties of the flow field, including polar flocks~\citep{Menzel:MolPhys:2018}, dynamical clusters~\citep{Liverpool:PRE:2017, Gompper:SoftMatter:2018}, and chaotic flows~\citep{Graham:PRL:2005,Graham:JPCM:2009,Qi:CommunPhys:2022}. However, due to the complexity and specific details of these swimmer models and how their confinement is treated computationally, a full analysis of the underlying causes behind the observed collective phenomena and their connection to system geometry is challenging. In this work, we seek to remedy this difficulty and instead use a minimal model designed to investigate the combined effects on collective motion from geometric restriction of the swimmer dynamics and hydrodynamic screening by the boundaries. The swimmers are restricted to move in a 2-dimensional sheet in the mid-plane between two solid, no-slip walls (Fig.~\ref{fig:1}a) separated by a fluid layer of height $h$, with each swimmer acting with a pair of equal and opposite point forces on the fluid. We assume that the confinement length $h$ is significantly larger than the microswimmer dimensions, so that friction between the swimmer body and the confining walls can be neglected~\citep{Bartolo:PRL:2013,Saintillan:PRE:2014}. In the limit $h \rightarrow \infty$, we recover a $2d$ sheet of microswimmers interacting via $3d$ flow fields, coinciding with the system studied by mean-field kinetic theory in~\citep{Skultety:JFM:2024}. The assumption that the swimmer positions and orientations are perfectly restricted to the midplane is clearly a simplification, introduced as a well-controlled way of isolating two important effects occurring exclusively for microswimmers swimming near boundaries, namely (\emph{i}) effective in-plane compressibility and (\emph{ii}) hydrodynamic screening of dipolar flows. While a sheet of microswimmers cannot be \emph{a priori} expected to remain stable in the absence of external forces~\citep{Yeomans:PRL:2022}, in experimental realisations this restriction of the swimmer dynamics is instead maintaned \emph{via} additional physical effects such as gravity, surface tension, or activity, that ensure that the microswimmers are restricted to move near the boundary~\citep{Stone:PNAS:2018,Theurkauff:PRL:2012,Fakhri:Nature:2022}. 

For pushers, we find that the nature of collective motion changes from one dominated by large-scale flows with correlation lengths comparable to the system size as observed in $3d$~\citep{Stenhammar1,Bardfalvy1} into motion dominated by small-scale vortices comparable in size to that of the individual swimmer. This behaviour is largely insensitive to the film thickness $h$, and is thus an effect of the $2d$ restriction of the swimmer dynamics rather than of the hydrodynamic screening caused by the walls. For pullers, we observe a discontinuous transition into dense puller clusters with an aster-like structure, corresponding to a predicted density instability for pullers restricted to a $2d$ sheet in a $3d$ fluid. For $h \rightarrow \infty$, the observed onset density $n^c_{2d}$ coincides with the analytically predicted one. $n^c_{2d}$ furthermore increases significantly with decreasing $h$, showing that hydrodynamic screening due to confinement acts to partially suppress this long-wavelength density instability. In summary, our results show that the geometric effect of restricting the swimmers' motion to a 2-dimensional subspace of the $3d$ bulk system is dominant in determining the dynamical state of the system, while the effect of hydrodynamic screening due to the presence of a boundary~\citep{Graham:PRL:2005,Graham:JPCM:2009} is subdominant. 

\begin{figure}[h!]
\center
\includegraphics[width=8.5cm]{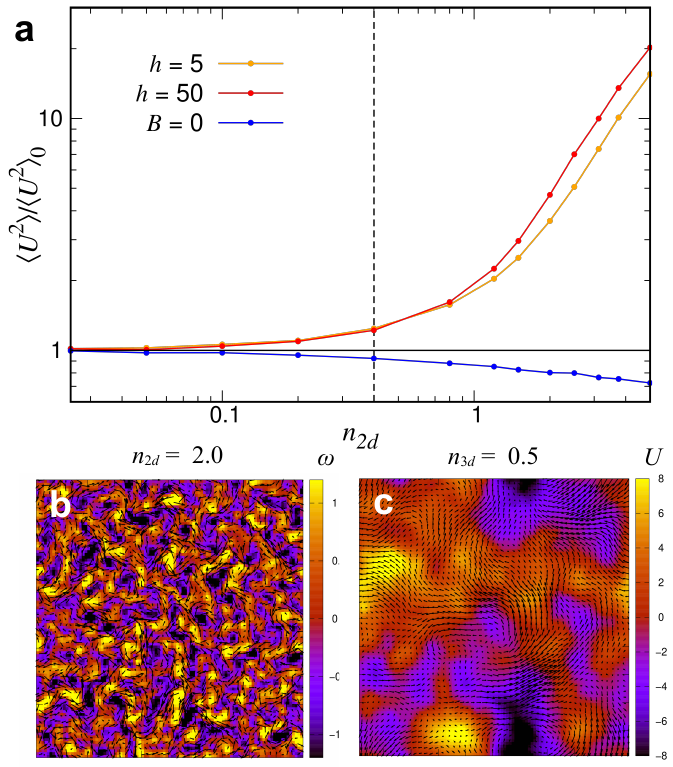}
\caption{\textbf{Collective motion in a layer of pushers.} (a) Fluid velocity variance $\langle U^2 \rangle$ for pushers normalised by the corresponding noninteracting value $\langle U^2 \rangle _0$ for two values of $h$, as indicated. Blue symbols show results for spherical swimmers with $B = 0$, and the vertical dashed line indicates the predicted onset density for $B = 1$ from Eq.~\eqref{eq:n_c_push}. (b,c) Snapshots of the fluid velocity field for $2d$ confined (b) and bulk $3d$ (c) pusher suspensions deep inside the collective motion regime, corresponding to $n_{2d} = 2.0$ and $n_{3d} = 0.5$~\citep{Bardfalvy1}, respectively. The quasi-$2d$ and $3d$ simulation boxes have the same lateral dimensions $H = 100$, and the former uses $h = 50$. The black arrows show the in-plane fluid velocity and the colours the vorticity (quasi-$2d$) and out-of plane velocity ($3d$), respectively.}\label{fig:2}
\end{figure}

\section*{Results and Discussion} \label{sec:results}

\subsection*{Model description} \label{sec:model}
We consider $N$ microswimmers swimming in a 2-dimensional plane centered between two parallel, flat walls separated by a distance $h$ in a simulation box with a quadratic base of side length $H$ (see Fig.~\ref{fig:1}a), yielding a 2-dimensional number density $n_{2d} = N/H^2$. The swimmer positions and orientations are restricted to the mid-plane of the simulation box ($z=h/2$) and periodic boundary conditions are applied in the $x$ and $y$ directions, while no-slip boundary conditions are applied in the $z$ direction. Each swimmer exerts equal and opposite forces $\pm \FF$ separated by a distance $l$ on the fluid, yielding a dipole strength $\kappa = \pm Fl/\mu$, where $F=|\FF|$ and $\mu$ is the dynamic viscosity of the suspending fluid; we use the convention that the microswimmer is considered a pusher if $\kappa > 0$ and a puller if $\kappa < 0$. The position $\rr_i$ and orientation $\pp_i$ of swimmer $i$ are governed by the equations of motion~\cite{Stenhammar1,Saintillan:PoF:2008,Nash2}:
    \begin{align}
    &\dot{\rr}_i=v_s\pp_i + \UU(\rr_i), \label{eq:rdot} \\[5mm]
    &\dot{\pp}_i=(\mathbb{I}-\pp_i\pp_i) \cdot [ \mathbb{W} + B \mathbb{E} ] \cdot \pp_i. \label{eq:pdot}
    \end{align}
Here, $v_s$ is the (constant) swimming speed, $\UU(\rr_i)$ is the fluid velocity evaluated at the position of swimmer $i$, $\mathbb{W}$ and $\mathbb{E}$ are respectively the  vorticity and rate-of-strain tensors, and $\mathbb{I}$ is the unit tensor. In addition to reorientation by the fluid, the microswimmers also change direction by random reorientation (``tumbling'') with Poisson-distributed frequency $\lambda$. Jeffery's equation~\eqref{eq:pdot} describes the angular dynamics of an elongated particle with shape parameter $B$ in a shear flow, and, unless otherwise stated, we consider the limit $B=1$, corresponding to a needle-like swimmer of infinite aspect ratio. We solve the model using the particle-based lattice-Boltzmann (LB) method described previously~\citep{Nash2,Bardfalvy1}; further details of the model parametrisation and units are presented in the Methods section.

\begin{figure*}[!htbp]
\center
\includegraphics[width=17cm]{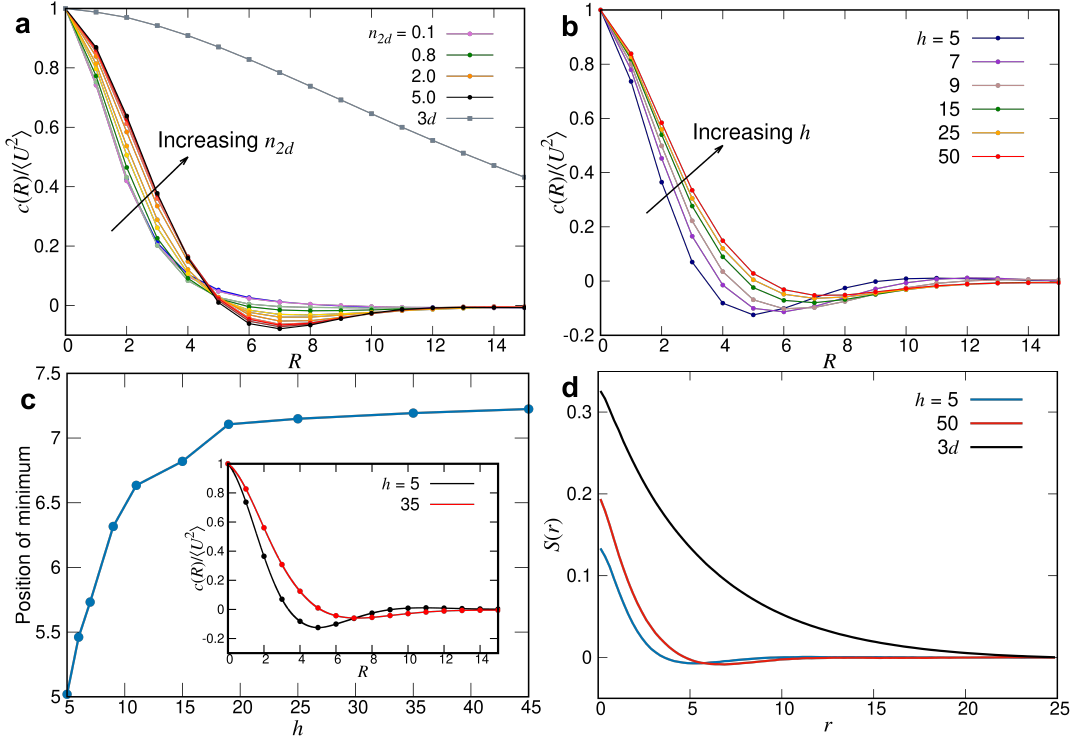}
\caption{\textbf{Spatial correlations in a layer of pushers are short-ranged.} (a,b) Normalised spatial correlation functions $c(R)$ of the fluid velocity in two dimensions for pusher suspensions for (a) $h = 50$ and varying density $n_{2d}$, and (b) for varying box dimensions $h$ and $n_{2d}= 5.0$. The gray line/square symbols in panel (a) shows the corresponding $c(R)$ in a $3d$ bulk suspension in the turbulent regime ($n_{3d} = 0.5$)~\citep{Bardfalvy1}. (c) Position of the primary minimum in $c(R)$ as a function of $h$, obtained by a cubic splines fitting to the data in (b) (inset). (d) Nematic order parameter $S(r)$ of pushers for two different film heights $h$ at $n_{2d} = 5.0$. The black line shows the corresponding result in the same $3d$ suspension as in (a).}\label{fig:3}
\end{figure*}

\subsection*{Pusher suspensions}

We begin by considering pusher suspensions, which are well-known to exhibit a transition to bacterial turbulence in unbounded suspensions. In Fig.~\ref{fig:2}a, we show the average of the fluid velocity variance $\langle U^2 \rangle / \langle U^2 \rangle_0$ for pusher suspensions as a function of the swimmer density, with $\langle U^2 \rangle_0$ being the corresponding quantity for a suspension of noninteracting swimmers, \emph{i.e.}, with all terms containing $\UU$ in the equations of motion \eqref{eq:rdot}--\eqref{eq:pdot} turned off. We can see that, for both film thicknesses studied ($h = 5$ and $h = 50$), $\langle U^2 \rangle / \langle U^2 \rangle_0$ starts to increase significantly above unity around $n_{2d} = 0.1$, indicating collective motion. Just as in unbounded, $3d$ pusher suspensions~\cite{Stenhammar1,Bardfalvy1}, this collective motion is driven by mutual reorientations: When changing the rotational dynamics~\eqref{eq:pdot}, to that corresponding to spherical swimmers with $B = 0$ (blue line in Fig.~\ref{fig:2}a), the average fluid velocity variance decreases dramatically, indicating a suppression of collective motion. In Fig.~\ref{fig:2}b, we furthermore show a snapshot of the fluid velocity and vorticity fields for a pusher suspension deep in the turbulent regime ($n_{2d} = 2.0$), together with the corresponding velocity field in an unbounded $3d$ suspension (Fig.~\ref{fig:2}c). Even though these two snapshots show systems at similar distances from the respective onset densities ($n_{2d}/n^c_{2d} \approx 5.0$ and $n_{3d}/n^c_{3d} \approx 4.7$, respectively), it is clear that the collective motion in the microswimmer sheet is dominated by much smaller-scale flow structures than in the bulk $3d$ system, as further emphasised in Supplementary Movies 1 and 2, respectively showing collective motion in $2d$ and $3d$. 

\begin{figure}[h!]
\center
\includegraphics[width=8.5cm]{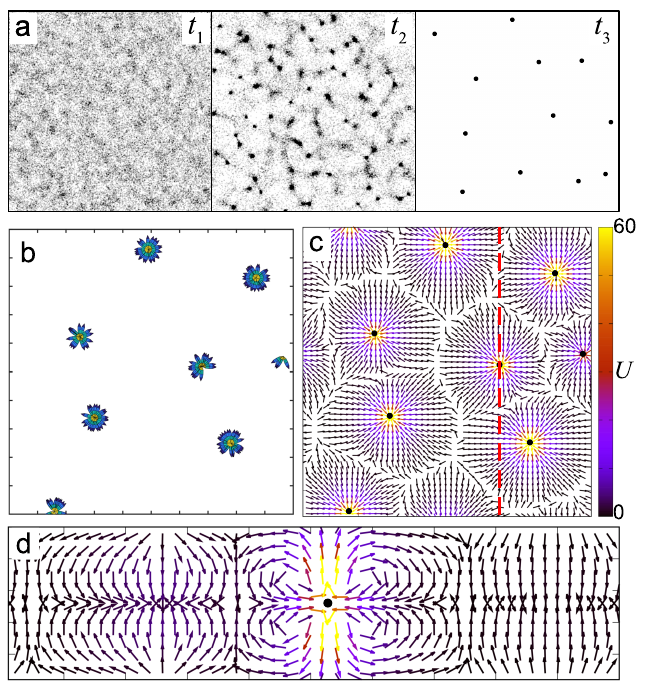}
\caption{\textbf{Clustering in a layer of pullers.}  (a) Snapshots of cluster formation in a sheet of pullers at three consecutive times after initialisation from a random configuration at $h = 25$ and $n_{2d} = 5.0$. (b) Visualisation of the microswimmer orientation after clustering. The swimmers are visualized as rods, with the swimming direction in yellow; note that, for visualization purposes, only a small subset of the swimmers in each cluster are shown. (cd) Velocity field in (c) and perpendicular to (d) the plane of the swimmers for the same configuration as in (b). The location of the $yz$-cut is indicated by the red dashed line in (c). Note that puller clusters act as effective sinks in the swimmer plane, while expelling fluid into the $z$-direction.}\label{fig:4}
\end{figure}

In order to quantify this observed difference, in Fig.~\ref{fig:3} we plot the spatial correlations of the fluid velocity, $c(R) = \langle \UU (\mathbf{0}) \cdot \UU(\RR) \rangle$, where the angular brackets denote a temporal and spatial average over all points for which $|\RR| = R$. Figure~\ref{fig:3}a shows the evolution of $c(R)$ as the density $n_{2d}$ is increased in a sheet of microswimmers confined between two walls with $h = 50$. As the pusher density is increased, the correlation function develops a clear minimum, indicative of vortical flows with a selected lengthscale. The minimum in $c(R)$ occurs at a lengthscale of $\sim 7$ times the swimmer size, \emph{i.e.}, significantly smaller than the system dimensions ($H = 100$, $h = 50$) but comparable to the microswimmer run-length. This is qualitatively different from the corresponding function in $3d$ (black line in Fig.~\ref{fig:3}a), where $c(R)$ instead decays algebraically over lengthscales comparable to the system size~\citep{Bardfalvy1}. In Fig.~\ref{fig:3}b, we furthermore show $c(R)$ deep in the turbulent regime calculated for different film thicknesses varying from $h=5$ to $h=50$. For all values of $h$, it is clear that our qualitative observations made for $h = 50$ are essentially unchanged: as the film thickness is decreased, the minimum of $c(R)$ is only slightly shifted towards shorter distances, showing that the typical lengthscale of collective motion is largely unaffected by the presence of in-plane hydrodynamic screening. This is confirmed by the  analysis in Fig.~\ref{fig:3}c, which shows the position of the primary minimum as a function of $h$: the position of the minimum quickly saturates at a value of $\sim 7$ times the swimmer length, comparable to the bacterial run length but much smaller than the system dimensions, around $h \sim 20$. This shows that the short-ranged nature of collective motion is not an effect of hydrodynamic screening, but rather an effect of the restricted dynamics of the swimmers, as we discuss further below. In Fig.~\ref{fig:3}d, we finally show the two-dimensional swimmer-swimmer nematic order parameter $S(r)$, defined by $S(r) = 2\langle \cos^2 \theta \rangle_{r} - 1$, where $\theta$ is the angle between the orientations of two swimmers separated by a distance $r$. Clearly, for both extremes of the film thickness ($h=5$ and $h=50$), $S(r)$ is qualitatively very similar to $c(R)$, showing that the fluid correlations are driven by hydrodynamically induced, short-ranged nematic ordering between pushers. The nematic order is less pronounced and decays faster than in $3d$, in spite of the larger tendency for close encounters that lead to nematic alignment between swimmers in $2d$ compared to in $3d$. 

Further insight can be gained by analysing the above results in light of our recent analytical predictions from kinetic theory~\cite{Skultety:JFM:2024}, where we showed that $3d$ pushers restricted to a $2d$ plane exhibit a hydrodynamic instability due to the mutual, in-plane reorientation induced by HIs. Differently from $2d$ or $3d$ unbounded bulk suspensions, the most unstable wavenumber $k_c$ corresponds to the smallest wavelength available in the system, \emph{i.e.}, $k_c \rightarrow \infty$. While the choice of this small-scale cutoff is not \emph{a priori} obvious, we chose the corresponding lengthscale, $l_c = 2\pi/k_c$, to correspond to the average swimmer-swimmer separation, below which the continuum description of the system breaks down. This assumption leads to the following expression for the critical density $n^c_{2d}$:
	\begin{equation}\label{eq:n_c_push}
		n_{2d}^c = \frac{4}{\pi} \left( \frac{\lambda}{B\kappa} \right)^{2/3} + 2 \sqrt{2} \frac{v_s}{B\kappa},
	\end{equation}
where the first term on the right-hand-side is due to the short-range cutoff introduced by the finite value of $k_c$. Note that, unlike the bulk $3d$ instability criterion~\eqref{eq:nc_3d}, the criterion for a sheet of pushers explicitly depends on the swimming speed $v_s$, in addition to the dipole strength $\kappa$ and tumbling rate $\lambda$. Similar to the bulk $3d$ case, the critical density diverges as $B \rightarrow 0$, illustrating that the instability is driven by swimmer reorientation and therefore is absent for spherical swimmers, as illustrated numerically in Fig.~\ref{fig:2}a. Inserting our simulation parameters yields $n_{2d}^c = 0.40$, which is shown by the vertical line in Fig.~\ref{fig:2}a. While the very gradual increase to collective motion makes it difficult to quantitatively verify the instability criterion in~\eqref{eq:n_c_push}, the predicted critical density coincides well with the density region where the velocity variance starts increasing significantly above its noninteracting value (note the double logarithmic scale in Fig.~\ref{fig:2}a). The observations made in Fig.~\ref{fig:3} furthermore makes it clear that the short-wavelength nature of the pusher instability manifests itself as a qualitatively different form of collective motion than in bulk $3d$ systems. Since the analysis in~\citep{Skultety:JFM:2024} corresponds to the limit $h \rightarrow \infty$, it furthermore supports the notion that this is not an effect of hydrodynamic screening of the in-plane flows, but rather of restriction of the microswimmer dynamics to a $2d$ subspace of the $3d$ fluid. 

\subsection*{Puller suspensions}

We now turn to the case of puller suspensions in the same quasi-$2d$ geometry, where we observe a distinctly different form of collective behaviour. For large enough densities $n^c_{2d}$, a sudden destabilisation occurs in the system, whereby localised, dense microswimmer clusters form (see Fig.~\ref{fig:4}a and Supplementary Movies 3 and 4). As shown in Fig.~\ref{fig:4}b, the internal structure of these clusters is aster-like, with pullers pointing their heads towards the centre of each cluster. In the plane of the swimmers, this leads to a sink-like flow field around each cluster (Fig.~\ref{fig:4}c), structurally similar to an array of +1 defects, separated by lines of low in-plane velocities where the fluid recirculates. The sink-like flows attract additional swimmers and eventually leads to cluster coalescence, although the late-stage coarsening is too slow for us to definitively assess the nature of the steady state. The flow field in the $z$ direction (Fig.~\ref{fig:4}d) points out of the swimmer plane above and below each cluster in order to satisfy the overall incompressibility of the fluid. Unlike the pusher case, this instability is fully driven by mutual swimmer-swimmer \emph{advection} rather than orientation, and the clustering thus completely disappears when the advection term in the swimmer dynamics~\eqref{eq:rdot} is turned off. 

\begin{figure}[h!]
\center
\includegraphics[width=8.5cm]{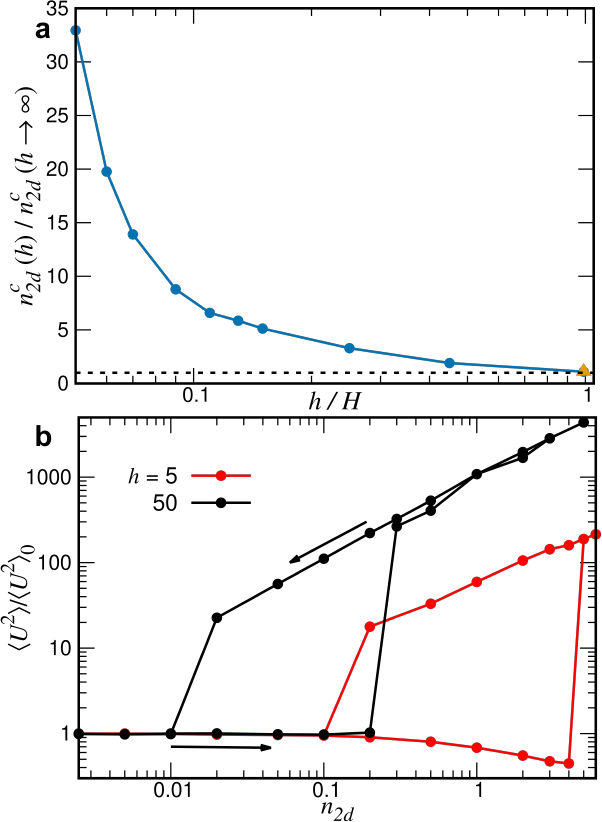}
\caption{\textbf{Clustering behaviour in a puller layer depends strongly on the confinement height.} (a) Measured onset density $n^c_{2d}$ for clustering in puller suspensions as a function of the film thickness $h$. As $h$ approaches the lateral system size $H$, $n^c_{2d}$ approaches the theoretical value from Eq.~\eqref{eq:nc_pull} (dashed line, $n^c_{2d}(h)/n^c_{2d}(h \rightarrow \infty) = 1$), valid in the $h \rightarrow \infty$ limit. The orange triangle shows the clustering density measured for $B = 0$, corresponding to spherical swimmers, illustrating that the instability is independent of swimmer shape. (b) Hysteresis curve for pullers, showing the normalized fluid velocity variance for two different film thicknesses while cycling the total density first upwards, then downwards, as indicated. The system density was increased (decreased) by adding (removing) swimmers in random positions and then letting the system reach steady state at the new density. Note the significant hysteresis effects due to the discontinuous nature of the transition. }\label{fig:5}
\end{figure}

The above observations are in line with the derived hydrodynamic instability for pullers in~\citep{Skultety:JFM:2024}. There, we showed that pullers confined to move in a $2d$ plane are unstable to density fluctuations, and that the instability occurs at the longest wavelength available to the system. In the case of a vertically unbounded fluid ($h \to \infty$), this wavelength corresponds to the lateral system size $H$, leading to a critical density $n^c_{2d}$ given by
\begin{equation}\label{eq:nc_pull}
n^c_{2d} = \frac{8\pi}{H} \frac{v_s^2}{\lambda |\kappa|}.
\end{equation}
This criterion shows that a sheet of pullers is unstable for \emph{any} density in the infinite-system limit $H,h \rightarrow \infty$. Unlike the pusher instability in Eq.~\eqref{eq:n_c_push}, $n_{2d}^c$ is independent of $B$, showing that it is indeed driven by interparticle advection rather than reorientation. To test the prediction of Eq.~\eqref{eq:nc_pull}, in Fig.~\ref{fig:5}a we plot the observed onset density $n^c_{2d}$ for clustering as a function of the film height $h$, where $n^c_{2d}$ is defined as the lowest density where a random initial configuration of microswimmers goes unstable towards clustering. Since the instability is a long-wavelength one, it is strongly affected by the hydrodynamic screening induced by confinement, which considerably increases $n^c_{2d}$ for small $h$. As $h$ is increased the onset density decreases, and for $h = H = 100$ reaches the theoretically predicted value from Eq.~\eqref{eq:nc_pull} (dashed line in Fig.~\ref{fig:5}a) both for needle-like ($B = 1$) and spherical ($B = 0$) swimmers. The strong $h$-dependence of the dynamics is also evident in the kinetics of the clustering and subsequent coarsening (see Supplementary Movies 3 and 4), which become significantly faster as $h$ is increased. In Fig.~\ref{fig:5}b we furthermore show hysteresis curves, where the total density of the system is gradually changed by first adding and then removing swimmers with respect to the previous configuration, and letting the system reach steady state at each new density. Both film thicknesses exhibit strong hysteresis effects, highlighting the discontinuous nature of the transition, in stark contrast with the collective motion in pusher suspensions where we observe no history dependent effects.  

The very high microswimmer densities observed in the point-like puller clusters after the instability are clearly not representative of actual microswimmer suspensions due to the absence of excluded volume interactions and near-field HIs. Thus, studying the internal dynamics of the clusters after the instability is not physically meaningful. Nevertheless, due to its long-wavelength nature and since the puller instability occurs at arbitrarily dilute concentrations ($n_{\mathrm{2d}}^{c} \rightarrow 0$ as $H \rightarrow \infty$), the far-field description remains relevant for describing the onset of clustering in real-world realisations of puller suspensions. We finally note that the occurrence of aster-like clusters in puller suspensions was previously observed by Li and Ardekani~\citep{Ardekani_PRL_2016}; these were however observed in bulk $2d$ suspensions where in-plane flow fields are fully incompressible, and must therefore be due to a different physical mechanism than those observed here. 

\section*{Conclusions} \label{sec:conclusions}

In this work, we have studied the combined effects of in-plane restriction and hydrodynamic screening on the collective motion of microswimmer suspensions, using a minimal dipolar swimmer model of biological microswimmers such as bacteria and algae. The swimmers were restricted to move in the mid-plane between two rigid, non-slip walls with varying separation $h$ to qualitatively mimic the effect on collective motion from confinement near a solid or fluid interface. While simplistic, this setup allows us to carefully test the combined effects of (\emph{i}) geometric restriction of the swimmer dynamics to a $2d$ plane, and (\emph{ii}) hydrodynamic screening due to the confinement by no-slip walls on the ensuing collective motion. Our results are compatible with the hydrodynamic instabilities in a sheet of microswimmers derived using mean-field kinetic theory in the unconfined ($h \rightarrow \infty$) limit~\citep{Skultety:JFM:2024} and supports the interpretation of these instabilites as indicators of the onset of collective motion. This connection is however not obvious \emph{a priori}: While the linear pusher instability in $3d$ has generically been interpreted as being equivalent to the onset of bacterial turbulence in unconfined systems~\citep{Saintillan1}, a general connection between such instabilities and the ensuing, strongly nonlinear states is not possible. Firstly, the pusher instability occurs at short scales, where it is unclear whether the coarse-graining procedure used to derive the kinetic theory is accurate. Secondly, the analytical model neglects the effect of hydrodynamic screening due to confinement, where our numerical results indicate that the observed collective behaviour is indeed qualitatively robust even in the presence of physical boundaries.

For a sheet of pushers, our results show that the short-wavelength hydrodynamic instability corresponds to a significantly different type of collective motion characterised by small-scale vortical flows compared to in $3d$ unbounded suspensions. In the latter case, the emerging lengthscales are much larger than the microswimmer size, and are expected to diverge close to the onset to collective motion~\citep{Bardfalvy1}. For the model parameters used here, chosen to mimic those of \emph{E. coli}, the critical density $n^c_{2d} \approx 0.4$ from Eq.~\eqref{eq:n_c_push} roughly corresponds to an experimental number density of $\SI{0.1}{\per\square\micro\meter}$, which is relatively high but not unphysical: assuming a film height of $h = \SI{5}{\micro\meter}$, it corresponds to a volume density of $\sim \SI{2d10}{\per\milli\litre}$, comparable to experimental \emph{E. coli} densities~\citep{Gachelin1}. The emerging picture of short-ranged collective motion in a microswimmer sheet naturally raises the question of whether the far-field hydrodynamic picture used here remains relevant for describing active turbulence in films or other geometrically restricted microswimmer systems. The fact that the instability in a sheet of pushers occurs at wavelengths comparable to the swimmer size means that the far-field HIs will likely be outcompeted by near-field HIs and nonhydrodynamic forces such as excluded volume. Since these interactions are more specific to each organism than the generic far-field flows, our results indicate that, differently from dilute $3d$ suspensions, quantitative predictions for confined systems are unlikely to be accessible using generic microswimmer models. 

For sheets of pullers, we see even more striking differences compared to 3-dimensional systems, where no collective behaviour is present for pullers. For pullers restricted to a $2d$ sheet, we observe a sharp transition to a clustered state, occurring at a density that decreases with $h$. The critical density for this advection-driven onset of clustering is in quantitative accordance with predictions from mean-field kinetic theory in the unconfined ($h \rightarrow \infty$) limit~\cite{Skultety:JFM:2024}, again validating our interpretation of this hydrodynamic instability. The density instability highlights a peculiar property of swimming near a confining wall: since swimmers can expel fluid into the third dimension, the in-plane flow fields are effectively compressible without violating the global incompressibility condition, leading to significantly enhanced density fluctuations compared to in unbounded microswimmer suspensions, where no density instabilities are present. 

To summarise, we have demonstrated a number of striking differences between collective motion in bulk and in microswimmers moving in a plane. Since our results take account for two generic aspects of microswimmers moving in confined geometries, they highlight that the details of both the sample geometry and swimmer dynamics need to be carefully considered when analysing experimental and computational results in active systems dominated by long-ranged interactions. 

\section*{Methods} 

We solve the model using the particle-based lattice-Boltzmann (LB) method described previously~\citep{Nash2,Bardfalvy1}. In LB units, measured in terms of the lattice spacing $\Delta x$ and timestep $\Delta t$, the swimmers have a length $l = 1$ and swim with a constant speed of $v_s = 10^{-3}$. We furthermore used $\lambda = 2\times 10^{-4}$, $F = 1.55 \times 10^{-3}$, and $\mu = 1/6$. In these units, we used a lateral system size of $H = 100$ while the film height was varied between $h = 5$ and $100$. The swimmers interact solely through their long-ranged flow fields, where the divergence is numerically regularised as described in~\citep{Nash2}, and possess no additional, non-hydrodynamic interactions. Although this is obviously a simplification compared to real systems, this choice allows a systematic study of a minimal model system with a limited number of free parameters, in a spirit similar to the Vicsek model of polar flocks~\citep{Vicsek1}. The parameters correspond to a reduced run length $v_s / (\lambda l) = 5$ and velocity field strength $F l \lambda^2/(\mu v_s^3) = 0.37$, which are comparable to the corresponding quantities in \emph{E. coli} suspensions frequently used in experiments on bacterial turbulence. All results are presented in terms of the (dimensional) $2d$ microswimmer density $n_{2d} = N / H^2$, while lengths and times are non-dimensionalised using the swimmer length $l$ and persistence time $l/v_s$, respectively. For comparison, we also include some results obtained from unbounded, $3d$ pusher suspensions, as described previously in~\citep{Bardfalvy1}. 

\section*{Data availability}
Data sets generated during the current study are available from the corresponding author on reasonable request. 

\section*{Code availability}
The lattice Boltzmann software used to generate data for the study is available upon request from the corresponding author. 

\section*{Acknowledgements} 
Henrik Nordanger is kindly acknowledged for help with the visualisation. JS is funded by the Swedish Research Council (grant ID 2019-03718) and the Crafoord foundation. AM acknowledges financial support from EPSRC (grant number EP/V048198/1). CN was supported by the ANR grant PSAM. The research was partially funded through an International Research Program (IRP) of the Institute National de Physique (INP) ``IFAM''. The computations were enabled by resources provided by LUNARC. For the purpose of open access, the authors have applied a Creative Commons Attribution (CC BY) licence to any Author Accepted Manuscript version arising from this submission. 

\section*{Author contributions}
DB, VS, CN, AM, and JS conceived and designed the project. DB and JS developed the computational methodology and performed the simulations and data analysis. DB and JS wrote the manuscript with input from the other authors. 

\section*{Competing interests}
The authors declare no competing interests. 

\bibliography{bibliography}

\end{document}